\documentstyle[11pt,mrs2001,epsfig]{article}
\newcommand{\mpc}{{$h^{-1}$ Mpc}}
\newcommand{\sqdeg}{{$\mbox{deg}^2$}}
\newcommand{\etal}{{\sl et al.\/}}
\def\ltsima{$\; \buildrel < \over \sim \;$}
\def\lsim{\lower.5ex\hbox{\ltsima}}
\def\gtsima{$\; \buildrel > \over \sim \;$}
\def\gsim{\lower.5ex\hbox{\gtsima}}

\begin{document}

\title{The Alignment Effect of Brightest Cluster Galaxies in the SDSS}

\author{R.S.J. Kim$^{1,2}$, J. Annis$^3$, M.A. Strauss$^1$, R.H. Lupton$^1$, N.A. Bahcall$^1$, J.E. Gunn$^1$, J.V. Kepner$^{1,4}$, M. Postman$^5$ for the SDSS collaboration}
\affil{$^1$Princeton University Observatory, Princeton, NJ 08544, USA}
\affil{$^2$Department of Physics and Astronomy, The Johns Hopkins University, 3701 San Martin Dr, Baltimore, MD 21218, USA}  
\affil{$^3$Fermilab, Batavia, IL 60510, USA}
\affil{$^4$MIT Lincoln Laboratory, Lexington, MA 02420, USA}
\affil{$^5$Space Telescope Science Institute, 3700 San Martin Dr., Baltimore, MD 21218, USA}

\begin{abstract}
One of the most vital observational clues for unraveling the origin of
Brightest Cluster Galaxies (BCG) is the observed alignment of the BCGs
with their host cluster and its surroundings.  We have examined the
BCG-cluster alignment effect, using clusters of galaxies detected from
the Sloan Digital Sky Survey (SDSS).  We find that the BCGs are
preferentially aligned with the principal axis of their hosts, to a
much higher redshift ($z \gsim 0.3$) than probed by previous studies
($z \lsim 0.1$).  The alignment effect strongly depends on the
magnitude difference of the BCG and the second and third brightest
cluster members: we find a strong alignment effect for the dominant
BCGs, while less dominant BCGs do not show any departure from random
alignment with respect to the cluster.  We therefore claim that the
alignment process originates from the same process that makes the BCG
grow dominant, be it direct mergers in the early stage of cluster
formation, or a later process that resembles the galactic cannibalism
scenario.  We do not find strong evidence for (or against) redshift
evolution between $0<z<0.45$, largely due to the insufficient sample
size ($< 200$ clusters).  However, we have developed a framework by
which we can examine many more clusters in an automated fashion for
the upcoming SDSS cluster catalogs, which will provide us with better
statistics for systematic investigations of the alignment with
redshift, richness and morphology of both the cluster and the BCG.
\end{abstract}

\section{Introduction}
There has been an increasing number of reports in the past two decades
of coherent orientation between galaxies, clusters and large-scale
structure (see Djorgovski~1986 for review). 
%The moden era of alignment
%studies starts with Rood \& Sastry (1972), who found a mild alignment
%of early type galaxies in Abell 2199 with the cluster. Further studies
%(Hawley \& Peebles~1975, Thompson~1976, Godwin, Metcalfe \&
%Peach~1983, Dekel~1985) have claimed various degrees of alignments (or
%non-alignments). Recent efforts to detect or predict intrinsic galaxy
%shape correlations on larger scales, mostly in the context of weak
%lensing studies (Heavens, Refregier \& Heymans~2000, Croft \&
%Metzler~2000, Pen, Lee \& Seljak~2000), have not yet found such signal
%that would challenge the current hierarchical structure formation
%theory (McKay \etal~2001).
In particular, Bingelli (1982) reported the two very significant
alignment signals, both of which were later nick-named the ``Bingelli
Effect''.  With 44 low redshift Abell clusters ($z<0.1$), he reported
a significant tendency for (i) neighboring clusters to {\it point
towards each other}, and (ii) the Brightest Cluster Galaxies (BCG) to
be aligned with their parent cluster orientations.  The first Bingelli
effect (cluster-cluster) was later re-investigated, and some confirmed
the effect (Rhee \& Katgert~1987, West~1989), while others did not
(Struble \& Peebles~1985).  On the other hand, evidence for the second
Bingelli effect (BCG-cluster) has been growing steadily over the last
two decades.  Such studies have involved rich Abell clusters (Struble
\& Peebles~1985), poor groups of galaxies (Fuller, West, \&
Bridges~1999) and a full three-dimensional analysis of the Virgo
cluster (West \& Blakeslee~2000).

However, all BCG-cluster alignment studies to date only probe the
phenomenon at low redshifts ($z\lsim 0.1$), where most of the Abell
clusters available for such studies live, typically including a
handful of clusters.  Studying the BCG alignment effect as a function
of redshift with a large sample of clusters is crucial to draw any
conclusions on the formation process of the BCGs and the origin of
their alignment effect.  In this paper, we examine the BCG alignment
effect in a sample of 300 clusters of detected by two or more methods
in the Sloan Digital Sky Survey (York \etal~2000; SDSS), in a redshift
range $0.04 \lsim z \lsim 0.5$.  In the process of this work, we develop a
framework with which we can analyze a much larger set of upcoming SDSS
cluster samples in an automated fashion; a complete study will be
presented in a forthcoming paper (Kim \etal~2001d).  Throughout this
paper, we use $H_0 = 70 \mbox{km s}^{-1}\mbox{Mpc}^{-1}$ and a
cosmology in which $\Omega_m = 0.3$ and $\Omega_{\Lambda} = 0.7$.

\section{Defining the Cluster Sample}

\subsection{Data}
The SDSS imaging data is taken with an imaging camera (Gunn
\etal~1998) on a wide-field 2.5m telescope, in five broad bands
($u,g,r,i,z$).  The point source magnitude limit is $r^* \approx 22.5$
at $1.5''$ seeing (see York \etal~2000, Stoughton \etal~2001 for more
details). The SDSS imaging runs 752 and 756 are equatorial scans that
are a part of the Early Data Release (Stoughton \etal~2001), and
amounts to 230 \sqdeg\/ of contiguous area. We use approximately half
of this area, 113 \sqdeg\/ enclosed by the coordinates $145^{\circ} <
\mbox{RA} < 190^{\circ}$ and $-1.25^{\circ} < \mbox{DEC} <
1.25^{\circ}$ for our present study.  The median seeing FWHM was 1.4''
in Run 752, and 1.3'' in Run 756.  We construct a galaxy catalog
to $r^*=21.5$ using star-galaxy classification by the SDSS photometric
pipeline ({\tt photo}; Lupton \etal~2001).
%All magnitudes quoted are in Petrosian (1976) quantities, and all colors
%are calculated from model magnitudes (See Paper II for more detailed
%description of the data and processing).

\subsection{Cluster Sample Selection}
\label{sec:sample}
We use three automated cluster finding algorithms for selecting our
sample of clusters in the SDSS: two Matched Filter (MF) algorithms,
the Hybrid MF (HMF; Kim \etal~2001b, hereafter Paper I), the Adaptive
MF (AMF; Kepner \etal~1999), and the MaxBCG algorithm of Annis
\etal~(2001). The MF techniques select clusters in 2D data by finding
peaks in a cluster likelihood map, generated by convolving the galaxy
distribution with a matched filter constructed from a model cluster
profile and luminosity function.  Both MF techniques are
outlined in Paper~I, and have been applied to the SDSS imaging data in
Kim \etal~(2001c).  The MaxBCG technique identifies most likely BCGs
based on their well studied characteristics in color-magnitude (c-m)
space, and then searches in their vicinities for galaxies with the
colors and magnitudes of likely cluster members (i.e., the
red-sequence), to determine the presence of a cluster.  Due to the
limited space provided, we shall refer to Paper I, Annis \etal~(2001),
for the details of these methods (see also Kim \etal~2001a). 

%The possible selection biases against clusters with unusual properties
%rising from the restrictive color-magnitude selection criteria of this
%method is not of important concern: recall that our goal is to
%investigate clusters with a prominent BCG to begin with, and {\it
%not}, for example, to study the Butcher-Oemler (1980) effect.
%To be more conservative, we use the results of the matched filters 

In order to minimize false positives and select a robust sample, we
take only those clusters that are detected by both the MaxBCG and
either of the MF methods. The MaxBCG selects clusters with very
restrictive color properties, but no constraint is put on the cluster
luminosity function or the density profile, while the MF algorithms
use the latter criteria exclusively; the two methods are practically
orthogonal in their selection criteria. Therefore, combining the two
makes our cluster selection especially robust. We obtain 300 clusters
with estimated redshifts out to $z \sim 0.55$ in 113 \sqdeg\/ (see Kim
\etal~2001a,d for details).  The MaxBCG is particularly useful since
it identifies the most likely BCG of each cluster, although not always
correctly; in order to flag these incorrect BCG identifications we start
by inspecting the 300 cluster candidates by eye. Visual inspection
effectively allows us to confirm the identity of the cluster
candidate, and to locate the BCG when it is dominant, being either a
giant elliptical or a cD with an extended envelope.  Each cluster
candidate was carefully examined, and was selected if the BCG was
visually unambiguous {\it and} was correctly located by the MaxBCG.
We also rejected clusters that had two or more dominant galaxies.
This eliminates half of our sample, with a total of 144 clusters
remaining.  We call this sample ``VC'' (Visually Confirmed).

We also perform an automated procedure that carries out a similar task
of filtering out clusters with likely spurious BCG detections.  We
inspect the area in c-m space bound with $(g-r)_{bcg}{}^{+0.2}_{-0.1}$
and $r_{bcg}{}^{+0}_{-1}$,
% $(g-r)_0 -0.15 < (g-r) <(g-r)_0 + 0.15$ and $r_0 -1 < r <r_0$, 
where $(g-r)_{bcg}$ and $r_{bcg}$ are the color and magnitude of the
detected BCG.  If there is any galaxy in the above area, within 0.5
\mpc\/ of the detected BCG, then we reject this cluster since there is
a good chance that this brighter galaxy is the true BCG.  This
eliminates 89 candidates out of the 300 in our initial sample; of
these, only 10 out of the 89 had been previously accepted as good
candidates by visual inspection. We call the resulting 211 cluster
sample ``MC'' (Machine Confirmed).

\subsection {The Degree of ``Dominance''}
\label{sec:dod}
The VC sample does not include clusters in which the BCG is not
dominant and therefore not readily identifiable.  However, a cluster
will always have a brightest member regardless of how dominant it is,
all of which are included in the MC sample as long as the BCG is
correctly identified.  We quantify this ``degree of dominance'' by
the magnitude difference between the BCG ($m_1$) and the
average of the second ($m_2$) {\it and} the third ($m_3$) brightest
member: $m_{(1-2,3)} \equiv (m_2+m_3)/2 - m_1$.  The average of $m_2$
and $m_3$ is slightly more robust to background contamination than is
$m_2$, it also deweights the fact that there can be two dominant
galaxies far more luminous than the rest of the cluster population.
%We determine $m_2$ and $m_3$ within 0.5 \mpc\/ of the
%BCG, and a narrow color range enclosing the red-sequence,
%$(g'-r')_{bcg}{}^{+0.1}_{-0.2}$, which keeps the background
%contaminations low in the redshift range of interest ($z\lsim0.5$).

%The resulting distribution of $m_{(1-2,3)}$ is slightly skewed as
%expected from the existence of a lower limit ($m_{(1-2,3)} = 0$) and
%the peak occurs roughly at $m_{(1-2,3)} = 0.65$. Below, we will use
%this value to divide the sample into two BCG subgroups, those that are
%dominant and less dominant.

\section{Method and Analysis -- Shape and Orientation}

The orientation and ellipticities of individual BCGs are taken
from the outputs of {\tt photo}: the position angle $\phi_{deV}$ and
axis ratio $\alpha_{deV}$ are the parameters of the best fit 2-D de
Vaucouleurs law model (Lupton \etal~2001). 
%These were chosen not only because the de
%Vaucouleurs model best describes the BCG profiles, but also because
%they are derived from model fits {\it convolved with the PSF}, which
%takes away any systematics of the PSF that would be reflected in the
%parameters otherwise.
The shape and orientation of the clusters are measured through a
process very similar to detecting clusters with the Voronoi
Tessellation Technique (VTT) which is detailed in Paper I (\S2.2).
The VTT is invoked reversely in order to retreive the cluster members
with which we calculate the position angle and ellipticity of the
cluster.  For each cluster, we apply c-m cuts on all galaxies around
the detected BCG (Paper I, Eq.~(9)-(11)), which selects a generous set
of galaxies whose colors and magnitudes resemble the cluster members.
We then apply a Voronoi Tessellation on these galaxies to select those
residing in the densest environments as a tracer for typical cluster
members: galaxies with $\delta \equiv (\rho - \langle \rho \rangle) /
\langle \rho \rangle > 3$ within 1\mpc\/ of the BCG,
where $\rho$ is the inverse of the Voronoi cell area.  We call the
final number of selected galaxies, $N_{cc}$.  We then calculate the
inverse-radius weighted moments of these galaxies using the BCG as the
center (${\bf x'} = {\bf x} - {\bf x_{bcg}}$), and obtain the Stokes
parameters whose relation to the position angle ($\phi$) and axis
ratio ($\alpha = b/a$) are as follows,
\begin{eqnarray}
Q &\equiv& {1 - \alpha \over 1+\alpha} \cos 2\phi ~=~ \langle {x'^2 \over r^2} \rangle - \langle {y'^2 \over r^2} \rangle = 2\langle {x'^2 \over r^2} \rangle - 1 \label{eq:q} \\
U &\equiv& {1 - \alpha \over 1+\alpha} \sin 2\phi ~=~ 2 \langle {x'y' \over r^2} \rangle \, .
\label{eq:u} 
\end{eqnarray}

\begin{figure}
%\plotone{cba14_c1v3.ps}{2in}
\plotone{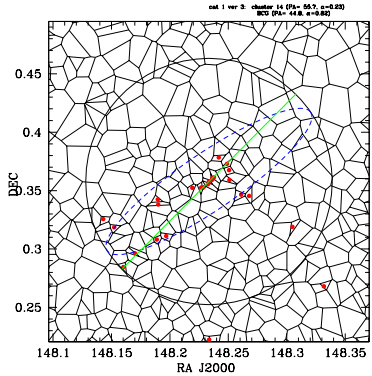}{2in}
\caption{Determining the cluster position angle. Voronoi Tessellation
is performed on c-m selected galaxies. The dots represent
galaxies with high density contrast, with which we compute
$\phi$ and $\alpha$. These parameters are represented by the dashed
ellipse around the BCG with its major axis normalized to 1\mpc\/
(solid circle). The solid line represents the PA of the BCG.  This
cluster is at z=0.24.
\label{fig:ex}}
\end{figure}

As with individual BCGs, when a cluster is close to being round, the
determination of the orientation becomes noisy and less meaningful.
Eqs.~(\ref{eq:q})~and~(\ref{eq:u}) show that the axis ratio $\alpha$
is directly related to the distance from the center of the $Q,U$ plane
($D \equiv \sqrt{Q^2 + U^2}$), $\alpha = (1-D)/(1+D)$.  When $D=0$ the
object is round, and becomes flatter as $D \rightarrow 1$.  Thus,
clusters with $(Q,U) = (0,0)$ less than $1\sigma$ away are consistent
with being round.  This translates to $\chi^2 \equiv (Q/\sigma_Q)^2 +
(U/\sigma_U)^2 \leq 2.3$ (Press \etal~1992), where $\sigma_Q,
\sigma_U$ are the shot noise errors on $Q, U$.  We eliminate all
clusters that either (a) have BCGs with $\alpha_{deV} > 0.9$, (b) are
{\it consistent with being round}, (c) have member galaxies $N_{cc} <
5$.  Each condition reduces the sample size about 20\%, leaving
altogether 88 clusters in the VC sample, and 115 clusters in the MC
sample (for the latter we also limit $z \leq 0.45$). We show an
example of a cluster with a well determined position angle and a
strongly aligned BCG in Figure~\ref{fig:ex}.

\section{Results}
\begin{figure}
\plottwo{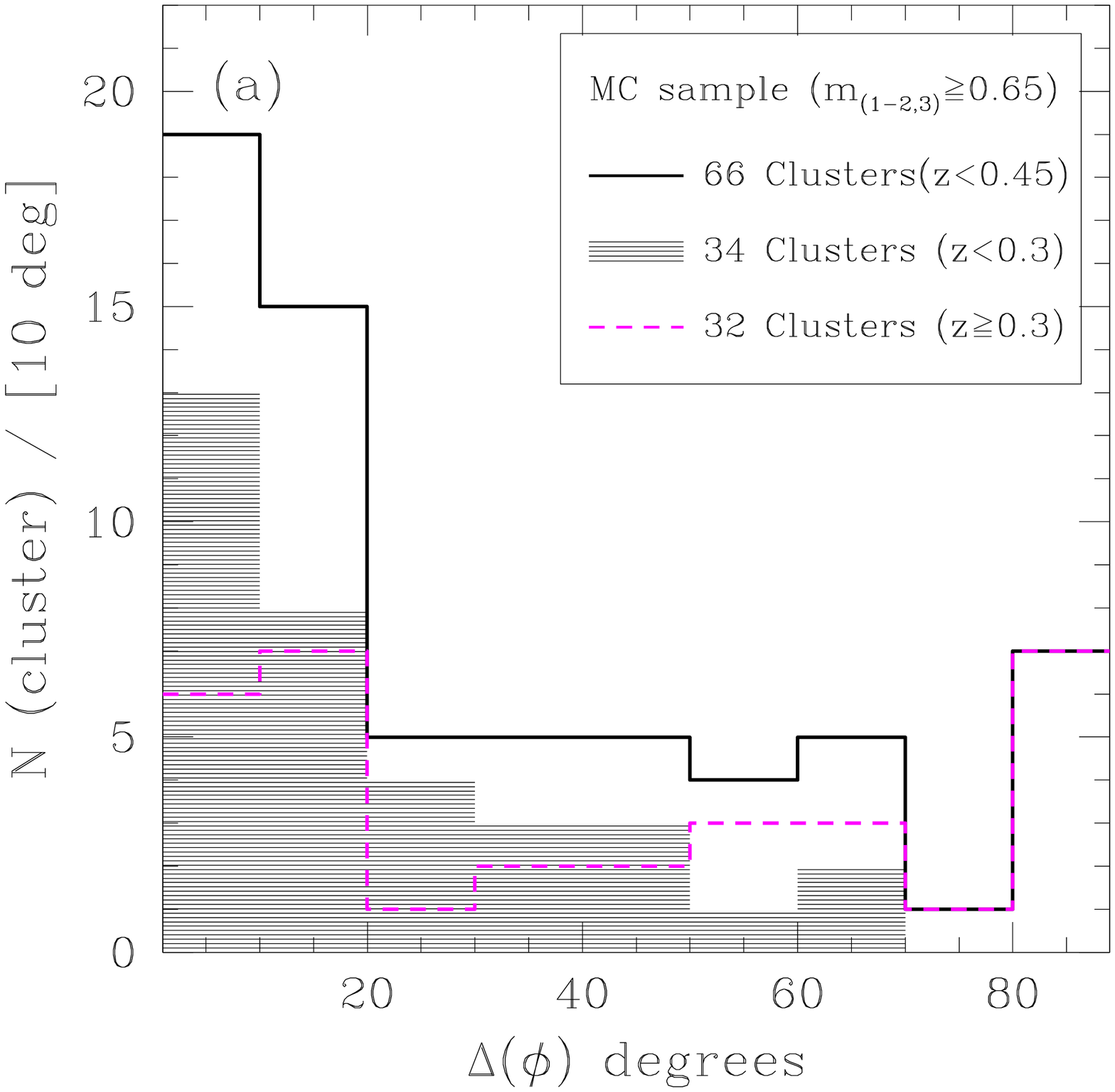}{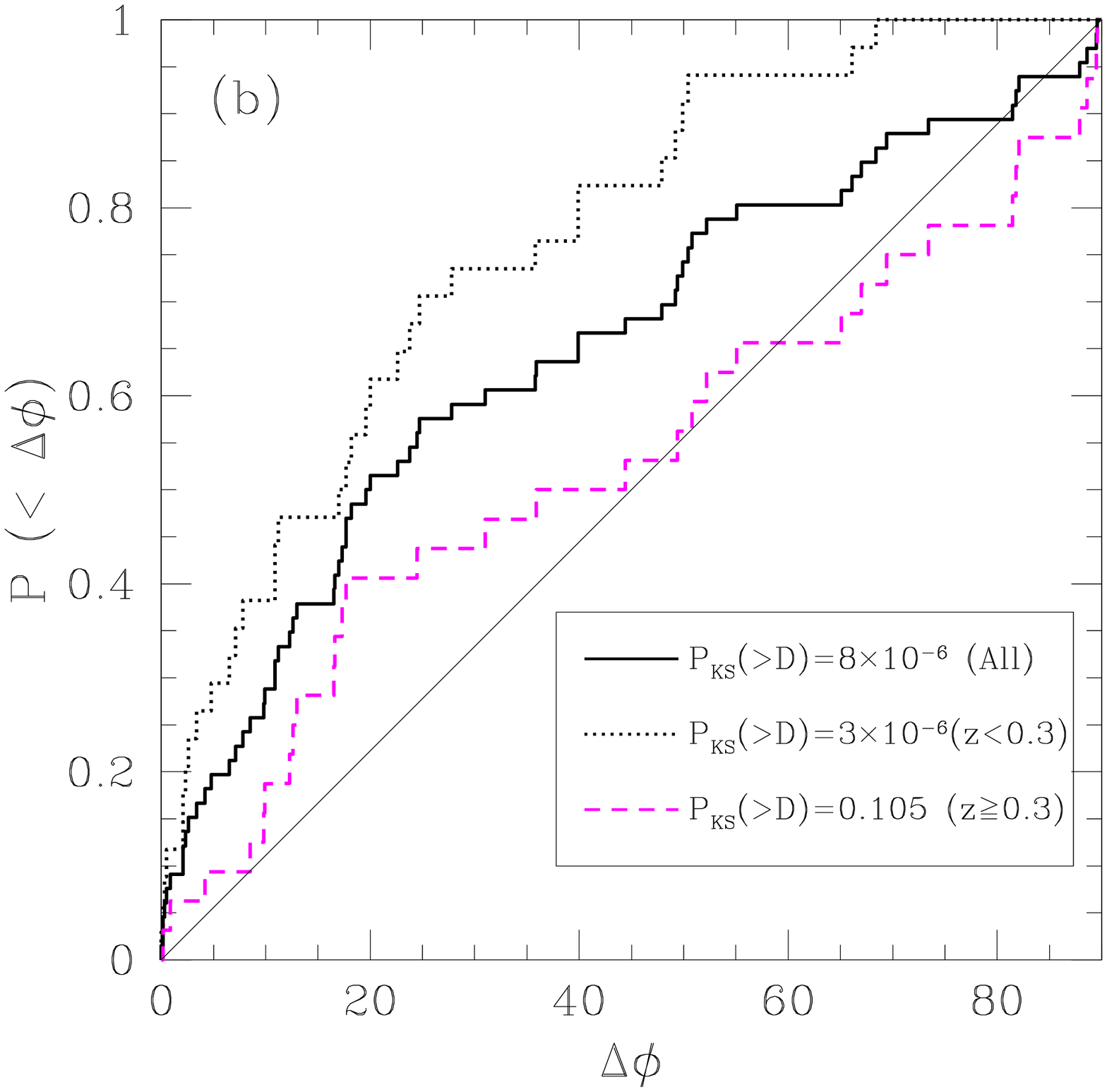}
\caption{ Histogram distribution of the difference in the position
angles of the BCG and the cluster, for 66 Machine Confirmed clusters
with dominant BCGs (MC sample I; solid-line). The same for 34 low $z$
clusters with $z <0.3$ (shaded) and 32 high $z$ clusters with $z\geq
0.3$ (dashed-line). (b) Cumulative probability distribution (CPD) of
the 66 MC sample I clusters (solid), the 34 low $z$ (dotted), and 32
high $z$ clusters (dashed).  All three distributions are significantly
different from a random distribution, showing an alignment signal at a
high confidence level.
\label{fig:resultmcg}}
\end{figure}

We divide the 115 clusters in the MC sample in two separate groups
according to their ``degree of dominance'' measured by the index
$m_{(1-2,3)}$ (\S\ref{sec:dod}): 66 clusters with $m_{(1-2,3)} \geq
0.65$ (MC sample I) and 49 clusters with $m_{(1-2,3)} < 0.65$ (MC
sample II).  We first present results for MC sample I, the cluster
sample with dominant BCGs.  Fig.~\ref{fig:resultmcg}a shows the
binned distribution of $\Delta\phi$ for all 66 MC sample I clusters
(solid-line).  
%We further split the sample into two redshift ranges,
%$z < 0.3$ and $z \geq 0.3$, the distribution of $\Delta\phi$ is shown
%also for the low $z$ (shaded) and the high $z$ (dashed-line)
%subsamples.  
Roughly half (34) of the total sample is confined within $\Delta\phi =
\pm 20^{\circ}$, where one would have expected only 14.7 clusters for
a random distribution.  This would imply that the number of clusters
within $\Delta\phi = \pm 20^{\circ}$, is more than $3\sigma$ above the
expected average in a random case.  We further split the sample into
two redshift ranges, $z < 0.3$ and $z \geq 0.3$, whose distribution of
$\Delta\phi$ is also shown in Fig.~\ref{fig:resultmcg}a.  Note the
striking signal especially at low $z$ (shaded).  When comparing these
two subgroups, there are hints towards a possible redshift evolution,
however, they may well be due to noisier data at higher $z$ and we
will not draw any conclusions regarding a redshift evolution, mainly
due to the small sample size.  The cumulative distributions (unbinned)
of $\Delta\phi$ are shown in Figure~\ref{fig:resultmcg}b.  The three
curves are for: all 66 (solid), 34 low $z$ (dotted) and 32 high $z$
(dashed) MC sample I clusters.  K-S tests show that the alignment
signals for the 66 clusters and the 34 low $z$ clusters are indeed
significant at more than $99.999\%$ confidence, while the high $z$
cluster sample show only a mild correlation at a 90\% significance.
Although the results for the VC sample with 88 clusters are not shown
in this proceeding, they are remarkably similar to the results for the
MC sample I (dominant BCGs) shown in Figure~\ref{fig:resultmcg}.  K-S
statistics indicate that the total VC sample and 46 low $z$ clusters,
both show strong alignments at 99.996\% and 99.96\% confidence levels,
and the 42 high $z$ clusters although to a lesser extent, at 95\%.
The fact that the VC sample results coincide with those of the MC
sample I confirms the validity of our automated method for choosing a
cluster sample with dominant BCGs.

\begin{figure}
\plottwo{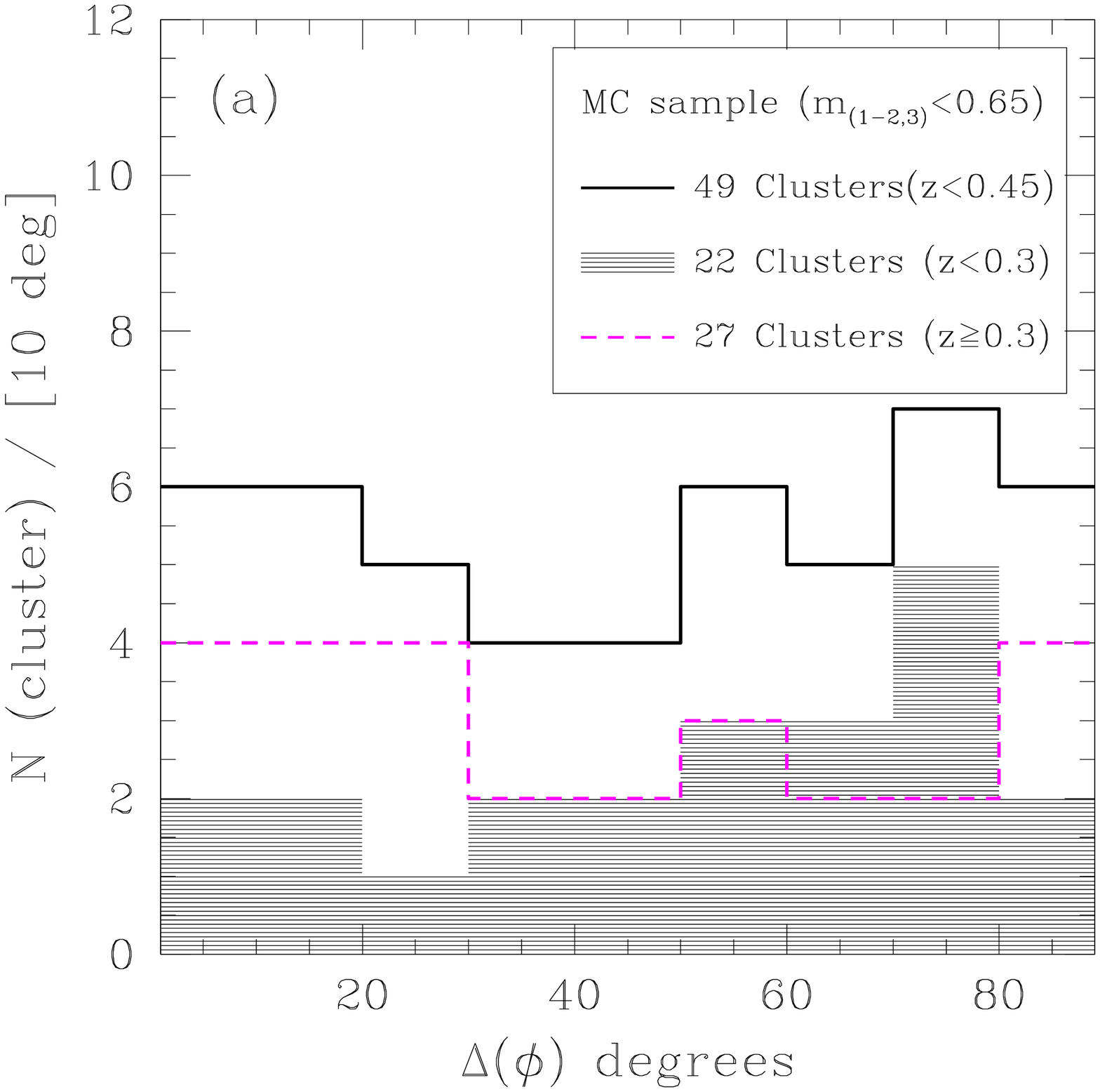}{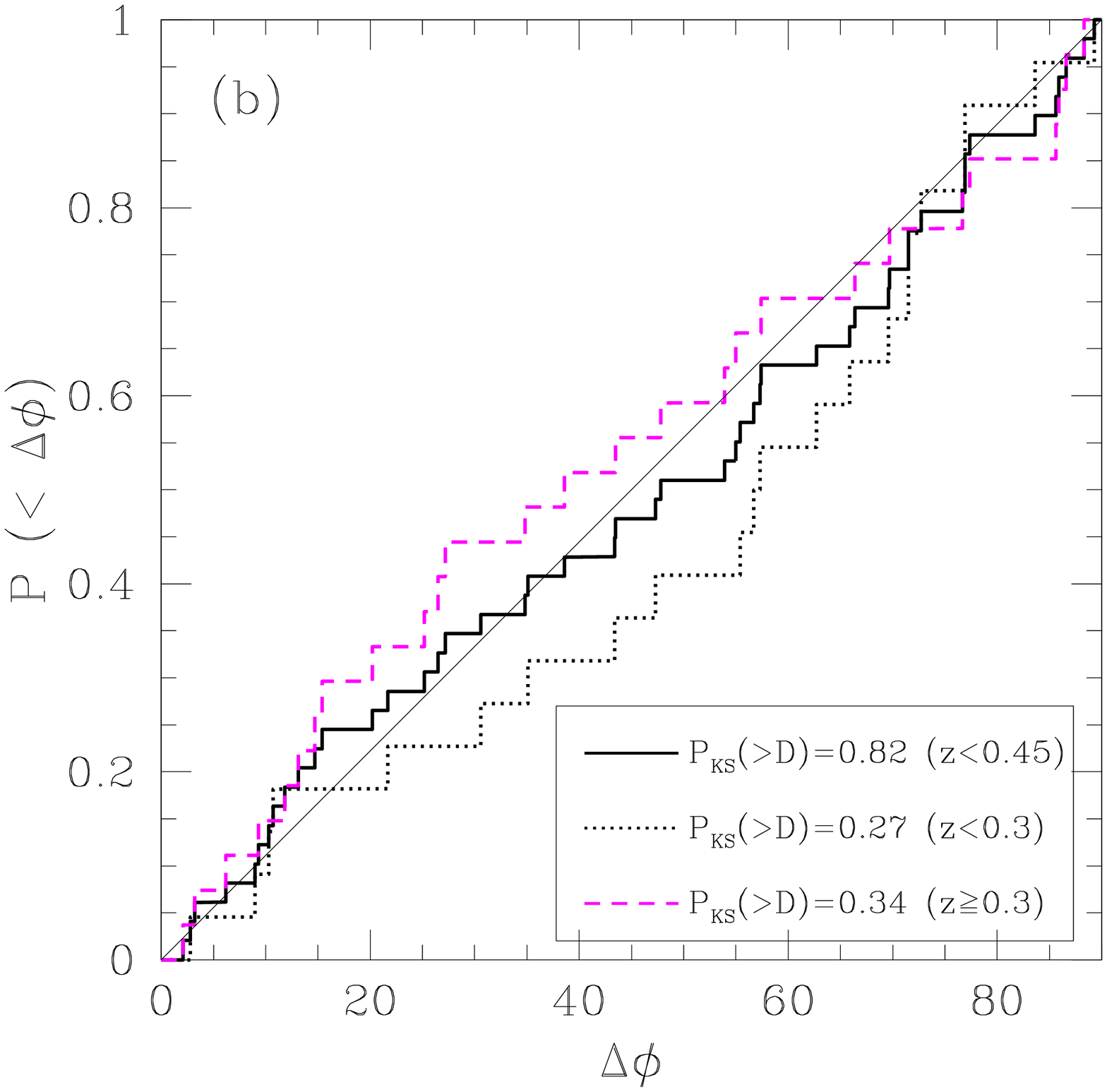}
\caption{Same as Fig.~\ref{fig:resultmcg}, for total 49 MC clusters
with less dominant BCGs selected by $m_{(1-2,3)} < 0.65$, and $z<0.45$
(MC sample II).  The distribution of these 49 clusters does not show
any departure from randomness whatsoever ($P_{KS}(>D) = 0.82$). The
low and high $z$ subsamples are also consistent with a random
distribution within $1\sigma$ ($P_{KS}(>D) = 0.27$ and $0.34$ for low
$z$ and high $z$, respectively).
\label{fig:resultmcl}}
\end{figure}

Finally, the 49 clusters with less dominant BCGs (MC sample II) show
an entirely different result.  Figure~\ref{fig:resultmcl} shows the
same as Figure~\ref{fig:resultmcg} for the MC sample II. The alignment
signals have totally vanished, and all three curves in
Figure~\ref{fig:resultmcl}b have distributions consistent with being
random.  In particular, the $\Delta\phi$ of all 49 clusters (solid) is
consistent with a random distribution at a 82\% significance level,
well above $1\sigma$ ($=32\%$). The $\Delta\phi$ of the low and high
$z$ subsamples are also consistent with a random distribution at 27\%
and 34\% significance ($ \sim 1\sigma$).  Such absence of the
alignment signal is quite remarkable when compared to the strong
signal seen for the dominant BCGs.

\section{Discussion}

Currently, the most favored theory for the origin of the BCG-cluster
alignment effect is the anisotropic infall of galaxies that merge to
produce BCGs that echo the preferred direction of cluster collapse
(Dubinsky~1998, Fuller, West \& Bridges~1999).  Prompted by
predictions of filamentary structures being a generic feature of
hierarchical structure formation models with Gaussian initial
conditions (Bond, Kofman, Pogosyan~1986, Weinberg \& Gunn~1990), this
picture portrays clusters forming through the infall of material along
the filaments, which would tend to align clusters with the nearby
large scale structure 
%(Bond~1987 Rhee \& Roos~1990, 
(van Haarlem \& van de Weygaert~1993).  Similarly, the formation of the
central BCG would have likely undergone anisotropic infall of galaxies
as well, assuming that they {\it are} merger remnants.  Indeed,
numerical simulations of galaxy collisions have successfully
reproduced a prolate de Vaucouleurs--like merger product aligned with
the initial collision trajectory, both with individual galaxy
collisions (White~1978) and cluster collapses in a cosmological
context (Dubinski~1998).

The core result of our work is: (i) we have confirmed the alignment
effect of BCGs with their parent clusters seen at low redshifts
($z<0.1$), to a much higher redshift ($z>0.3$), and (ii) only the
dominant BCGs show such alignments while clusters with less dominant
BCGs do not.  The first result supports the view that dominant
BCGs have relatively early origins, since the dominant BCGs in
clusters are aligned with their parent clusters already at a redshift
of $z \gsim 0.3$.  The second result, however, at face value, would
naturally seem to fit in the galactic cannibalism scenario (Ostriker
\& Tremaine~1975): dominant BCGs grow by accreting nearby bright
galaxies, making the magnitude difference between the BCGs and their
surroundings increase with time.  In this process the BCG becomes
aligned with the surrounding structure, which explains why less
dominant BCGs are {\it not} seen to be aligned with their
surroundings.  However, the late evolution scenario by galactic
cannibalism has well known drawbacks: the total accreted luminosity in
the cluster's lifetime falls short (by a factor of several) of the
current observed luminosities of typical cDs (Tremaine 1990).  
Early BCG formation during cluster formation
(consistent with hierarchical clustering), will be more efficient in
producing such dominant BCGs, through direct mergers of galaxies
(head-on collisions) in a preferred direction that can explain the
alignment effect.  However, this scenario does not explain the
existence of so many clusters with not so dominant BCGs at low
redshifts ($z<0.3$); either those clusters are yet in the process of
forming, or they have never had a chance to form a dominant BCG at the
early cluster formation stage and still never will.  The former seems
unlikely since the clusters we have selected are by definition those
with well evolved galaxies with uniformly red color (E/S0), and the
latter would imply that clusters have inhomogeneous origins and fates.

All these possibilities for the origin of BCGs and their alignment
effect with their surroundings can only be narrowed down effectively
by observing these effects with many more clusters in all possible
redshift ranges.  Examining the alignment effect or the evolution of
$m_{(1-2,3)}$ with cluster richness or BCG morphology can also provide
us with vital clues.  In our present study, we have established an
automated framework to investigate the alignment effect in many more
clusters ($>10,000$) in the SDSS and will soon be able to characterize
the redshift evolution of these effects for a valuable insight into
the origin of the brightest cluster galaxies.

We thank various people in the SDSS collaboration for their comments and 
support. RSJK acknowledges the support of NSF grants AST 98-02980, 
AST96-16901, AST-0071091, and NASA LTSA NAG5-3503.

\end{document}